\documentclass[12pt, letterpaper]{article}

\usepackage{amsmath,amssymb}
\usepackage{cite}
\usepackage{fancyhdr}
\usepackage[top=1in, bottom=1in, left=1in, right=1in]{geometry}
\usepackage{graphicx}
\usepackage{hyperref}

\numberwithin{equation}{section}
\date{}

\begin{document}
\title{{\rm\footnotesize \qquad \qquad \qquad \qquad \qquad \ \qquad \qquad \qquad \ \ \ \ \ \                      RUNHETC-2024-15
}\vskip.5in    Note on 't Hooft's Shock Wave Commutators From Near Horizon Conformal Field Theory}
\author{Tom Banks\\
NHETC and Department of Physics \\
Rutgers University, Piscataway, NJ 08854-8019\\
E-mail: \href{mailto:tibanks@ucsc.edu}{tibanks@ucsc.edu}
\\
\\
}

\maketitle
\thispagestyle{fancy} 

\begin{abstract}  We construct a finite model of 't Hooft's shock wave commutation relations from the ansatz\cite{Carlip}\cite{Solodukhin}\cite{BZ} that the quantum degrees of freedom in a causal diamond in a solution of Einstein's Equations are those of a (cut-off\cite{BZ}\cite{hilbertbundles} ) $1 + 1$ dimensional conformal field theory (CFT) with central charge related to the area of the diamond's holographic screen by equating Cardy's formula with the generalized Bekenstein-Hawking formula\cite{ted95}\cite{fsb}.  The particular CFT is an exactly marginal perturbation of free massless fermions, as motivated by the Holographic Space-Time\cite{hilbertbundles} (HST) program.  The momentum cutoff in $1 + 1$ dimensions and the Dirac eigenvalue cutoff on the transverse geometry, modify the 't Hooft relations when the area of the diamond is finite and remove all coincident point divergences in predictions derived from these commutation relations.  
\normalsize \noindent  \end{abstract}


\newpage
\tableofcontents
\vspace{1cm}

\vfill\eject
\section{Introduction}

 In 1990 't Hooft, extrapolating from an eikonal calculation of shock wave scattering he had done with Dray\cite{tHD}, postulated the commutation relations
 \begin{equation} [P^+ (\Omega), P^- (\Omega^{\prime})] = i(\Delta - R) (\Omega, \Omega^{\prime}) . \end{equation} Here $\Delta$ is the Laplacian on the holographic screen of the diamond, which in Minkowski or Anti-de Sitter space is its bifurcation surface, and $R$ is the scalar curvature.  $P^{\pm}$ are null momentum densities propagating on the boundaries of the diamond to the future/past of its {\it holographic screen}. For the scattering problem originally considered by 't Hooft the diamond is eventually taken to be the entire Penrose diamond of Minkowski space-time.  In subsequent years, these commutation relations have been derived directly from the Einstein-Hilbert Lagrangian in the near horizon limit where one focuses on small space-like distances from the boundaries of a causal diamond.  The time evolution parameters one uses in these derivations then resemble those defined by {\it modular flow}\cite{CHM} inside a diamond.  We will be particularly interested in half-sided modular flow, because it comes from the natural picture of building up a causal diamond in terms of a nested set of smaller causal diamonds (Figure 1).
 
 \begin{figure}[h]
\begin{center}
\includegraphics[width=01\linewidth]{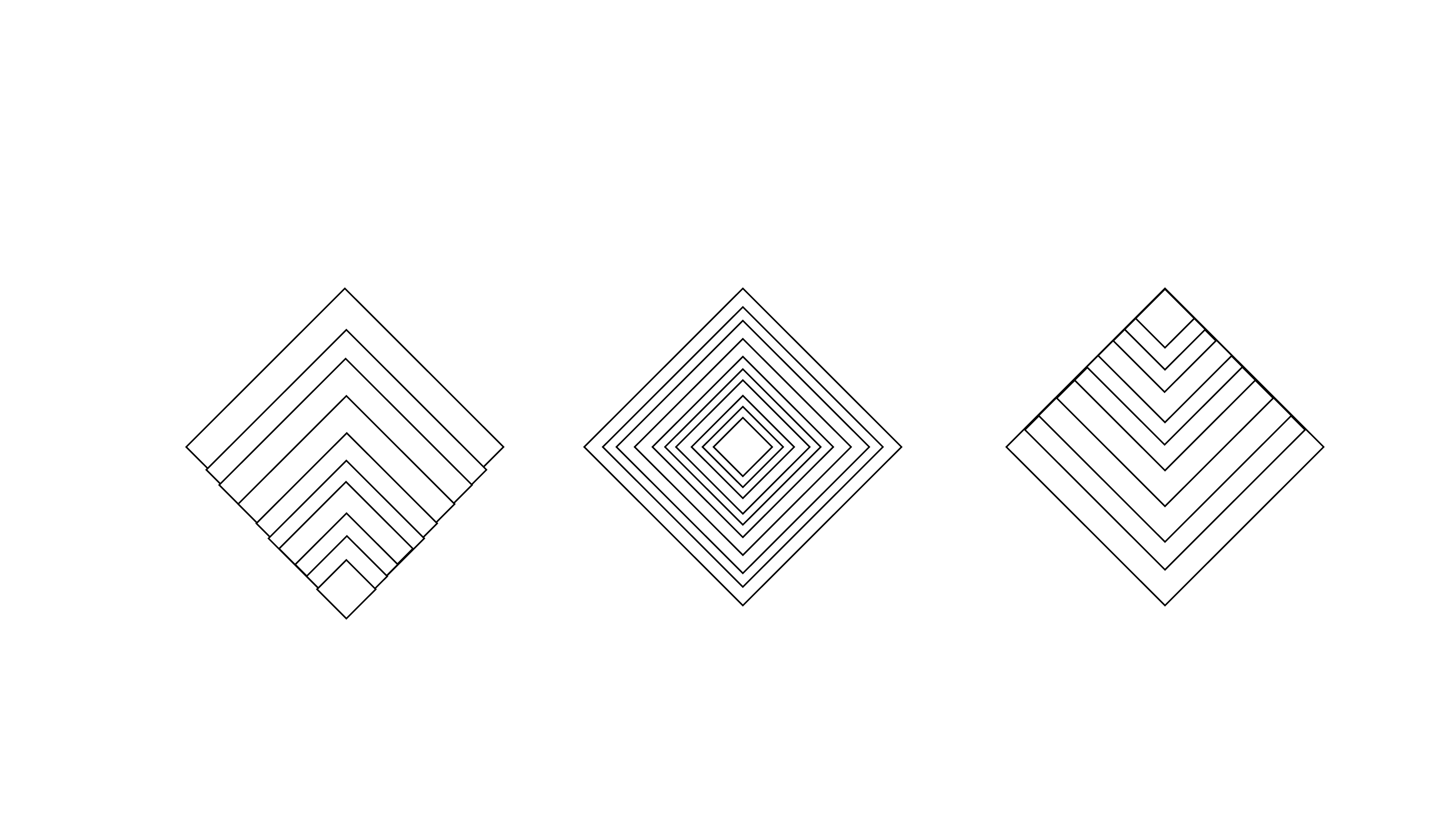}

\caption{Future directed, time symmetric, and past directed nested coverings of a causal diamond.} 
\label{fig:nestedcoversofadiamond}
\end{center}
\end{figure}

  If one believes\cite{ted95}\cite{fsb}, that once diamonds have large area, each Planck time step in such a nesting adds, for space-time dimension $d \geq 4$ a large number of q-bits to the system, there is no reason to expect collective variables like $P^{\pm}$ to be smooth functions of time.  They should have independent fluctuations in each Planck time interval, as first postulated in\cite{VZ1}.   Justification for this hypothesis in terms of a nested causal diamond picture was given in\cite{BZ} and\cite{tbwffluct}.   
 
 't Hooft's formula is singular at coinciding angles.  The commutator for the length variables $X^{\pm}$ conjugate to $P^{\pm}$ is less singular, but even in four space-time dimensions it has a logarithmic singularity.  This gives rise to bothersome ambiguities when one tries to relate these formulae to observable fluctuations in interferometer experiments\cite{tbwffluct}\cite{VZ3}.  A more microscopic approach is called for.
 
 In\cite{hilbertbundles} I proposed that the 't Hooft commutators be realized in terms of the Schwinger term of a $U(1)$ current algebra built out of fermions in the Carlip-Solodukhin CFT describing the physics of the causal diamond boundary.  The proposal did not quite work, but as we will see in this paper, a small but significant fix cures the problem encountered there.  This proposal provides a built in cut off for the 't Hooft commutators, whose only ambiguity comes from a choice of the $1 + 1$ dimensional momentum cut off on individual fermion fields.
 The angular singularity is regulated because the number of modes of the Dirac operator on the holographic screen of the diamond is equal to twice the central charge of the Carlip-Solodukhin CFT, which is $\frac{A}{4G_N}$.
 
 Let us remind the reader of the proposal of\cite{hilbertbundles}.  We adopt Jacobson's prescription\cite{ted95} that a solution of Einstein's equations describes the hydrodynamics of a quantum gravitational system.  Each causal diamond is a subsystem, the expectation value of whose modular Hamiltonian $K_{\diamond}$ is
 \begin{equation} \langle K_{\diamond} \rangle = \frac{A_{\diamond}}{4G_N}. \end{equation}
 The holographic screen of a diamond is the Riemannian $ d - 2$ fold of maximal volume in a null foliation of the diamond's boundary, and we abuse language by calling that volume the diamond's area.  The C-S CFT lives on a small interval of two dimensional Minkowski space transverse to the holographic screen, which we call the stretched horizon.  The C-S ansatz, is essentially that $K_{\diamond} = L_0$ , the Virasoro generator of the CFT.  We use conventions where the spatial extent of the stretched horizon is $[0,\pi]$.  In\cite{hilbertbundles} we proposed to realize the CFT in terms of Dirac fermion fields $\psi_n (t,z)$.   The number of fields is roughly proportional to the expectation value of the modular Hamiltonian.  These fields parametrize fluctuations of the geometry of the holoscreen around its background value.  The idea here is to use Connes' notion\cite{Connes} that the Dirac operator $D$ on a Riemannian geometry completely determines the geometry.  We expand solutions of the fluctuating Dirac equation  $D_f \Psi_f = 0$ in terms of eigenspinors of the background Dirac operator
 \begin{equation}  D^a_b \chi^b_n (\Omega ) = \lambda_n \chi^a_n (\Omega ) . \end{equation}
 \begin{equation} \Psi_f = \sum \psi_n (t, z) \chi_n (\Omega ) . \end{equation}  Here $a,b$ are spinor indices of $Spin (d - 2)$.  The fact that we allow only a finite number of $\psi_n$ means that the geometry of the finite area holoscreens is "fuzzy"\cite{tbjk}.  It is also quantum mechanical because the $\psi_n$ are fluctuating quantum fields with the C-S density matrix in the "empty diamond" state defined by the background geometry.  
 
 Bilinears in the fluctuating field
 \begin{equation} \bar{\Psi}_{f\ b} (t,z,\Omega)\gamma^m \Psi^a (t,z, \Omega ) , \end{equation} can, for finite values of the central charge, be thought of as a collection of $1 + 1$ dimensional Kac-Moody currents on the stretched horizon.  This statement is of course modified by the cutoff on $ 1 + 1$ dimensional momenta, but we will ignore that in this paper.  The justification for this will be outlined in the appendix, but the logic is basically that we're trying to reproduce a result derived from Einstein's equations, which are valid only in the limit of large causal diamonds. The C-S ansatz goes beyond Einstein's equations but it is still just a semi-classical conjecture.  As the discussion in the appendix makes clear, the cutoff $k_c$ on $1 + 1$ dimensional momenta of individual fermion fields is essentially a cutoff on the minimal size of causal diamond for which we believe the C-S argument.  It has to do with the scale at which corrections to Einstein's equations become important. String theory has taught us that this scale is model dependent.  In the real world, only experiment can teach us what it is.  
 
 As a matrix in spin space, the Kac-Moody currents transform as the tensor product of two spinor representations of Spin (d - 2).  For odd d, we will always choose the reducible representation that is invariant under space reflection.  Then that tensor product is the direct sum of all $p -$th rank anti-symmetric tensors, where $0\leq p \leq d - 2$.  Contracting these with the $ d - 2$ bein $e_m^A$ we obtain $p -$ form currents on the holoscreen, of all ranks, which are also Kac-Moody currents on the stretched horizon.  General Kac-Moody currents have an equal time algebra
 \begin{equation}   [J^0_A (z), J^1_B (y)] = i F_{ABC} J^1_C (y) \delta (z - y) - i {\rm tr} [j_A j_B] \partial_z \delta (z - y) . \end{equation} Here $j_{A,B} $ are the matrices of the ordinary Lie algebra from which the Kac-Moody algebra is constructed.  In order to obtain something resembling the 't Hooft 
commutators we want to project onto the Schwinger term only.  So we integrate $\int f_m (z) J^m (z) , $ with
\begin{equation} f_0 (0) f_1 (0) = 0 , \ \ \ \   f_0 (0) \partial_z f_1 (0) = - 1 . \end{equation}  In\cite{hilbertbundles} we argued that the time and space components of the $1 + 1$ dimensional currents were naturally associated with currents $P^{\mp}$ on the past and future stretched horizons of the causal diamond because the two light front directions are exchanged.  The two boundaries meet only on the holographic screen, and this corresponds to the equal time point in the Kac-Moody commutation relation.   It should be pointed out that this fits well with the discussion above about independent length fluctuations in each causal diamond in a nesting.  It does not make sense to think of $P^{\pm}$ as both being defined everywhere on both boundaries of the diamond.  Each of them is defined smoothly only on half of the diamond.  The 't Hooft CR make sense as commutation relations between $P^{-}$ at each point on the past boundary, with a {\it different} $P^{+}$ coming from sequential nested diamonds.   There's a complementary, time reversed, picture in which a smooth $P^{+}$ commutes with a sequence of $P^{-}$ variables from nested past oriented diamonds.

  Our new proposal is somewhat different than that in\cite{hilbertbundles} and motivated by the fact that the variables described there converge naturally to the Awada-Gibbons-Shaw generalization of the Bondi-Metzner-Sachs algebra so Supergravity\cite{ags}\cite{tbags} rather than bosonic gravity emerges from the formalism.
 $\Psi^a (z,\Omega)$ is a spinor, and the "matrices" which appear in the Kac-Moody algebra of Fermion bilinears are differential operators on the spinor bundle.   There does not appear to be an operator on the spinor bundle whose square has a Dirac trace equal to $\Delta - R$.  However, it is well known that, acting on the tensor product of two spinor bundles 
 \begin{equation}   D \otimes 1 + 1 \otimes D \rightarrow (d + d^{\dagger}) , \end{equation} and 
 \begin{equation} (d + d^{\dagger})^2 = (\Delta - R_W) . \end{equation}  In the first of these equations, the symbol $\rightarrow$ means that we use the equivalence between the tensor product of spinors and the complex of forms.  The operator on the right hand side is the sum of the exterior derivative and its Hodge dual.  In the second equation $\Delta $ is the Laplacian on the complex of forms and $R_W$ is the Weitzenboch curvature operation, which maps $p -$ forms to $p - $ forms by contraction with the Ricci tensor.  For $p = 0$ $R_W$ is just multiplication by the scalar curvature.  
 
 This motivates us to define
 \begin{equation} (P^{+})_b^a (\Omega )  = \int dz f_0 (z) [\bar{\Psi}_b (z, \Omega) \gamma^0 D^a_c \Psi^c (z, \Omega) + D_b^d \bar{\Psi}_d (z, \Omega) \gamma^0 \Psi^a (z, \Omega)] .  \end{equation} 
 
 \begin{equation} (P^{-})_b^a (\Omega )  = \int dz f_1 (z) [\bar{\Psi}_b (z, \Omega) \gamma^1 D^a_c \Psi^c (z, \Omega) + D_b^d \bar{\Psi}_d (z, \Omega) \gamma^0 \Psi^a (z, \Omega)] .  \end{equation} 
 Then, ignoring the fuzzification on the holoscreen,
 \begin{equation} [P^{+}_p (\Omega ), P^{-}_q (\Omega^{\prime} )]   = (\Delta  - R_W)_p \delta_p (\Omega - \Omega^{\prime}) \delta_{pq} .\end{equation}  In writing this equation, we've introduced the following simplified notation
 \begin{equation} (P^{\pm})_b^a = \sum P_p^{\pm} (\Gamma_p)_b^a , \end{equation} where $P_p$ is a $p -$ form coefficient and $\Gamma_p$ is the anti-symmetrized product of Dirac matrices, contracted with the $d - 2$ bein tangent vectors.  The equation then follows from our general Kac-Moody formula and from the trace orthogonality of anti-symmetrized Dirac matrices.  In writing this equation we've ignored the cutoff on the eigenvalues of the Dirac operator $D$.  The justification for this is that we are trying to reproduce a result that 't Hooft and others derived from the Einstein-Hilbert action, which is only valid in the limit of large diamonds.  
 
 We propose the following physical interpretation of this equation. We have argued\cite{tbags}\cite{hilbertbundles} that the Hilbert Bundles formalism predicts Supersymmetry in the limit of vanishing cosmological constant.  Supersymmetric theories of gravity generically have a variety of stable and unstable extended objects or $p\ - $ branes, as well as particles.  We would like to interpret the operators $P^{\pm}_P (\Omega) $ as the amount of null momentum carried by the piece of $p\ - $ brane world volume that happens to pass through the point $\Omega$ on the holoscreen.  General Relativity, which knows nothing of $p\ -$ branes, only picks up the $0$ brane piece of this.  
 
 Note however that while the $0 -$ brane piece of $P^{\pm}$ gets contributions from only the $1\ -$ form part of the fermion bilinear, the $0\ - $ form piece of the {\it fluctuations} is coming from $0\ - $ form bilinears.  So one cannot really say that we have identified the 't Hooft operators in the Hilbert bundle formalism.  Rather, they should be thought of as coarse grained averages, which capture a portion of the information encoded in the operators $P_p^{\pm}$ of the underlying quantum system.

  \section{Conclusions}
 
 Our analysis in this note more or less ignored the ultraviolet cutoffs that were an important part of the formalism of\cite{hilbertbundles}.  We have only used them somewhat implicitly in treating the Schwinger terms as properties of $1 + 1$ dimensional CFT.  That is we've treated the commutators of currents carefully, as if they were continuum $1 + 1$ dimensional fields, but treated the singularities on the holographic screen by naive canonical methods.  This is justified by the UV cutoff in the eigenvalue of the transverse Dirac operator $D$, which also regulates the ambiguities that were encountered in both\cite{VZ3} and\cite{tbwffluct}.  In principle, we should also take into account the fact that the individual fermion fields have a cutoff momentum spectrum $k_c$.  We believe that this is much less significant when the number of fermion fields is very large, as is the case for macroscopic causal diamonds.  
 
 What remains to be done in order to have a completely predictive model of the interferometer measurements is to relate these calculations more carefully to the time correlations that are actually measured.  Let's recall the various steps involved.  To go from the commutation relations to the size of the fluctuations in a single diamond, one needs to know the state of the system in the diamond.  Time reversal invariance\cite{tbwffluct} proves that it is the minimal uncertainty state for the canonically conjugate variables $P^{\pm}$.  The crucial point of the argument for the large size of the fluctuations was first intuited in\cite{VZ1}.  A slightly more detailed argument, in terms of the nested diamond picture and the fact that many new q-bits are added to a large diamond in a single Planck time step, was given in\cite{BZ}.  This is reinforced by the fact that $P^{\pm}$ cannot both be viewed as continuous variables in any consistent way of quantizing the system.  
 
 The most difficult step is to go from fluctuations in a single diamond to unequal time correlations in two diamonds shifted by a relative time $\tau$.  For correlations between modular Hamiltonians at different times, a moderately convincing argument was given in\cite{tbwffluct}, but this argument needs some work for the arrival time correlation function of direct experimental interest\footnote{At least in the opinion of the present author.}.

\vskip.3in
\begin{center}
{\bf Acknowledgments }
\end{center}
 The work of T.B. was supported by the Department of Energy under grant DE-SC0010008. Rutgers Project 833012.   
 \section{Appendix}
 
 The Carlip-Solodukhin ansatz for the density matrix of a large area causal diamond is
 \begin{equation} \rho = \frac{e^{- L_0}}{{\rm Tr }\ e^{- L_0}} , \end{equation} where $L_0$ is the Virasoro generator of a CFT on the interval $[0,\pi]$, with large central charge.  By assumption, the expectation value of the modular Hamiltonian is dominated by the regime of eigenvalues where the Cardy formula is valid.  Thus
 \begin{equation} {\rm Tr }\ e^{- L_0} = \int_0^{E_c} 
 e^{\sqrt{2\pi c E/6}} e^{-E} \end{equation}  

 For large $c$ this integral is dominated by the saddle point
 \begin{equation} E* = 2\pi c / 6  . \end {equation}  To leading order in $c$, the value of all quantities computed are independent of the upper cutoff $E_c$ as long as the saddle point is within the range of integration.  Note that in the original Carlip and Solodukhin papers, Cardy's formula was evaluated at a Virasoro eigenvalue determined by the solution of the classical hydrodynamic equations of the CFT.  
 
 The modular Hamiltonian $K = L_0 + {\rm ln} ({\rm Tr }\ e^{- L_0})$ .  At large $c$ this is just evaluated at the saddle point, and the second term is sub-leading.  This leads to the equation
 \begin{equation} \frac{2\pi c}{6} = \frac{A}{4G_N} , \end{equation} which is valid up to logarithmic corrections in $c$.  
 
 We can also understand the physical meaning of the $1 + 1$ dimensional cutoff $k_c $ on individual fermion fields.  From the theoretical point of view, the cutoff $k_c$ on individual fermion momenta starts to become significant when $c$ is not large.  In order to continue to use the C-S ansatz, we still have to assume that our system has large entropy and the Cardy formula is approximately valid.  We can no longer use the  saddle point approximation of the spectral sum.  This means that $|k_c|$ has to be as large as $15 - 20$.  The expectation value of $K$ with therefore be $ \gg 1$.   This means that the area of the smallest diamond to which we can apply the C-S ansatz is much larger than the Planck area.  Of course this is completely reasonable, since the C-S ansatz was based on the Einstein-Hilbert action and neglected short wavelength corrections to hydrodynamics of quantum gravity.  We may expect that in theoretical models, the actual value of the cutoff will depend on the moduli of the string model.   In the real world it can be determined only by experiment.

 \vfill\eject



\begin{thebibliography}{99}
 \bibitem{Carlip} S.~Carlip, ``Black hole entropy from conformal field theory in any dimension,''
Phys. Rev. Lett. \textbf{82}, 2828-2831 (1999)
doi:10.1103/PhysRevLett.82.2828
[arXiv:hep-th/9812013 [hep-th]].
\bibitem{Solodukhin} S.~N.~Solodukhin, ``Conformal description of horizon's states,''
Phys. Lett. B \textbf{454}, 213-222 (1999)
doi:10.1016/S0370-2693(99)00398-6
[arXiv:hep-th/9812056 [hep-th]].
 \bibitem{BZ}T.~Banks and K.~M.~Zurek, ``Conformal description of near-horizon vacuum states,''
Phys. Rev. D \textbf{104}, no.12, 126026 (2021)
doi:10.1103/PhysRevD.104.126026
[arXiv:2108.04806 [hep-th]].
 \bibitem{hilbertbundles} T.~Banks,``Hilbert Bundles and Holographic Space-time Models,''
[arXiv:2306.07038 [hep-th]].
 \bibitem{tHD} T.~Dray and G.~'t Hooft, ``The Gravitational Effect of Colliding Planar Shells of Matter,''
Class. Quant. Grav. \textbf{3}, 825-840 (1986)
doi:10.1088/0264-9381/3/5/013
 \bibitem{CHM} H.~Casini, M.~Huerta and R.~C.~Myers,
``Towards a derivation of holographic entanglement entropy,''
JHEP \textbf{05}, 036 (2011)
doi:10.1007/JHEP05(2011)036
[arXiv:1102.0440 [hep-th]].
 \bibitem{ted95} T.~Jacobson, ``Thermodynamics of space-time: The Einstein equation of state,''
Phys. Rev. Lett. \textbf{75}, 1260-1263 (1995)
doi:10.1103/PhysRevLett.75.1260
[arXiv:gr-qc/9504004 [gr-qc]].
 \bibitem{fsb} W.~Fischler and L.~Susskind,``Holography and cosmology,''
[arXiv:hep-th/9806039 [hep-th]];
R. Bousso, ``A Covariant entropy conjecture,''
JHEP \textbf{07}, 004 (1999)
doi:10.1088/1126-6708/1999/07/004
[arXiv:hep-th/9905177 [hep-th]];
R.~Bousso,
``Holography in general space-times,''
JHEP \textbf{06}, 028 (1999)
doi:10.1088/1126-6708/1999/06/028
[arXiv:hep-th/9906022 [hep-th]];
R.~Bousso,
``The Holographic principle for general backgrounds,''
Class. Quant. Grav. \textbf{17}, 997-1005 (2000)
doi:10.1088/0264-9381/17/5/309
[arXiv:hep-th/9911002 [hep-th]].
 \bibitem{VZ1} E.~P.~Verlinde and K.~M.~Zurek, ``Observational signatures of quantum gravity in interferometers,''
Phys. Lett. B \textbf{822}, 136663 (2021)
doi:10.1016/j.physletb.2021.136663
[arXiv:1902.08207 [gr-qc]].
 \bibitem{VZ3} E.~Verlinde and K.~M.~Zurek, ``Modular fluctuations from shockwave geometries,''
Phys. Rev. D \textbf{106}, no.10, 106011 (2022)
doi:10.1103/PhysRevD.106.106011
[arXiv:2208.01059 [hep-th]].
 \bibitem{tbwffluct} T.~Banks and W.~Fischler, ``Fluctuations and Correlations in Causal Diamonds,''
[arXiv:2311.18049 [hep-th]].
 \bibitem{tbjk} T.~Banks and J.~Kehayias,
``Fuzzy Geometry via the Spinor Bundle, with Applications to Holographic Space-time and Matrix Theory,''
Phys. Rev. D \textbf{84}, 086008 (2011)
doi:10.1103/PhysRevD.84.086008
[arXiv:1106.1179 [hep-th]];
\bibitem{Connes} A.~Connes, {\it Noncommutative Geometry}, Academic Press, San Diego, ???


 \bibitem{ags} M.~A.~Awada, G.~W.~Gibbons and W.~T.~Shaw, ``CONFORMAL SUPERGRAVITY, TWISTORS AND THE SUPER BMS GROUP,''
Annals Phys. \textbf{171}, 52 (1986)
doi:10.1016/S0003-4916(86)80023-9

 \bibitem{tbags} T.~Banks, ``Current Algebra on the Conformal Boundary and the Variables of Quantum Gravity,''
[arXiv:1511.01147 [hep-th]];
T.~Banks, ``The Super BMS Algebra, Scattering and Holography,''
[arXiv:1403.3420 [hep-th]].

 
 \end{thebibliography}


\end{document}